\documentstyle[12pt]{article}
\input epsf.tex

\newenvironment{frontpage}{%
        \setcounter{page}{0}
        \pagestyle{empty}}
        {\newpage
        \setcounter{page}{1}}

\newcommand{\preprint}[1]{%
        {\rightline{#1}}}

\renewcommand{\title}[1]{%
        {\begin{center}
        \Large\bf #1
        \end{center}}
        \vskip .3in}

\renewcommand{\author}[1]{%
        {\begin{center}
        #1
        \end{center}}}

\newcommand{\institution}[1]{\vspace{-1.7em}\vspace{0pt}
        {\begin{center}
        \it #1
        \end{center}}}

\renewcommand{\abstract}[1]{%
        \begin{center}%
        {\vspace{1em}\vspace{0pt}\bf Abstract}%
        \end{center}%
        \noindent #1}

\renewcommand{\date}[1]{%
        \begin{center}%
        #1%
        \end{center}}

\makeatletter

\renewcommand{\section}{\@startsection{section}{1}{\z@}%
    {-3.5ex plus -1ex minus -.2ex}%
    {2.3ex plus.2ex}%
    {\centering\normalsize\bf}}

\renewcommand{\subsection}{\@startsection{subsection}{2}{\z@}%
        {-3.25ex plus -1ex minus -.2ex}%
        {1.5ex plus .2ex}%
        {\centering\normalsize\it}}

\makeatother

\newenvironment{references}{%
        \newpage
        \begin{thebibliography}{99}%
        \frenchspacing}
        {\end{thebibliography}}

\makeatletter
        \@addtoreset{equation}{section}%
\makeatother

\newcommand{\beq}{\begin{eqnarray}}
\newcommand{\eeq}{\end{eqnarray}}

\def\endignore{}
\def\ignore #1\endignore{}

\newcommand{\too}{\longrightarrow}

\newcommand{\twi}{\widetilde}
\newcommand{\mybar}[1]%
        {\kern 0.8pt\overline{\kern -0.8pt#1\kern -0.8pt}\kern 0.8pt}
\newcommand{\sla}[1]%
        {\raise.15ex\hbox{$/$}\kern-.57em #1}
\newcommand{\roughly}[1]%
        {\mathrel{\raise.3ex\hbox{$#1$\kern-.75em\lower1ex\hbox{$\sim$}}}}

\newcommand{\drawsquare}[2]{\hbox{%
\rule{#2pt}{#1pt}\hskip-#2pt
\rule{#1pt}{#2pt}\hskip-#1pt
\rule[#1pt]{#1pt}{#2pt}}\rule[#1pt]{#2pt}{#2pt}\hskip-#2pt
\rule{#2pt}{#1pt}}

\newcommand{\Yfund}{\raisebox{-.5pt}{\drawsquare{6.5}{0.4}}}

\newcommand{\nop}[1]{:\kern-.3em#1\kern-.3em:}

\newcommand{\Group}[2]{{\hbox{{\sl #1}($#2$)}}}
\newcommand{\U}[1]{\Group{U\kern0.05em}{#1}}
\newcommand{\SU}[1]{\Group{SU\kern0.1em}{#1}}
\newcommand{\SL}[1]{\Group{SL\kern0.05em}{#1}}
\newcommand{\Sp}[1]{\Group{Sp\kern0.05em}{#1}}
\newcommand{\SO}[1]{\Group{SO\kern0.1em}{#1}}

\newcommand{\jref}[4]{{\it #1} {\bf #2}, #3 (#4)}

\newcommand{\NPB}[3]{\jref{Nucl.\ Phys.}{B#1}{#2}{#3}}

\newcommand{\PLB}[3]{\jref{Phys.\ Lett.}{#1B}{#2}{#3}}

\newcommand{\PRD}[3]{\jref{Phys.\ Rev.}{D#1}{#2}{#3}}

\def\twi{\widetilde}
\def\dtwi#1{\skew1\widetilde{\widetilde{#1}}}

\def\vbr{\vphantom{\sqrt{F_e^i}}}

\begin{document}

\begin{frontpage}

\preprint{}
\preprint{BUHEP-96-35}
\preprint{YCTP-P20-96}


\title{A Diagramatic Analysis of Duality in Supersymmetric Gauge Theories}

\author{Nick Evans}

\institution{Sloane Physics Laboratory,
Yale University,
New Haven, CT 06520, USA\\ 
and Department of Physics,
Boston University,
Boston, MA 02215, USA\\
\vskip.1in
{\tt nick@physics.bu.edu}}

\author{Martin Schmaltz}

\institution{Department of Physics,
Boston University,
Boston, MA 02215, USA\\
\vskip.1in
{\tt schmaltz@abel.bu.edu}}

\vskip.4in

\date{August, 1996}

\abstract{
We introduce a diagramatic notation for supersymmetric gauge theories. 
The notation is a tool for exploring duality and helps to present
the field content of more complicated models in a simple visual way.
We introduce the notation with a few examples from the literature.
The power of the formalism allows us to  
study new models with gauge group $(SU(N))^k$ and their  duals.
Amongst these are models which, contrary to a naive analysis,
possess no conformal phase.}

\end{frontpage}

\section{Introduction}
Following the initial work of Seiberg on supersymmetric QCD (SQCD) 
\cite{sqcd2} many examples
of duality in supersymmetric
$N=1$ field theories have been discovered. More recently, various groups have studied
more complicated models with several gauge and global symmetries and large 
numbers of fields \cite{manyduals,pop}. The commonly used notation showing tables
of fields and representations often proves to be very cumbersome.
We propose a diagramatical notation that shows how the fields
of the theory under consideration transforms under gauge and global symmetries.

The notation is based on a notation introduced by Georgi \cite{howard}
for the study of non-supersymmetric strongly coupled gauge theories (also used in the models
of \cite{othermoo}). 
In supersymmetric theories the exact results on the infrared spectrum of strongly
coupled theories
are particularly easily displayed using our ``Duality Diagrams''. Seiberg's
dualities have a particularly 
transparent realization in this notation. A simple set of ``Duality Rules''
which would be
amenable to implementation on a computer enable one to derive dualities of more complicated
product group theories. 
The proposed notation can be used for $SU$, $SP$, and $SO$ theories with
matter in the fundamental and tensor representations. We will concentrate 
mainly on $SU$ theories with fundamental matter in this paper.
In section 2, we introduce our notation using 
SQCD as an illustrative example. The next simplest model involves 
two $SU$ gauge groups and has been previously studied in \cite{pop}.
We reanalyze this model in section 2.2 as an example of the diagramatic
notation.

In  section 4, we use our notation to study  new models with matter transforming under
an $(SU(N))^k$ gauge group. When all possible Yukawa interactions
consistent with the gauge symmetry are included the model has an
$SU(F)^k$ global symmetry.
We find a dual description in terms of an $SU((F-N))^k$ gauge theory and a similar superpotential. 
For $F \ge 2N$ the 
original theory is infrared free, whereas for $F \le 2N$ the dual is free in the infrared. 
The theory therefore has no conformal phase. At the
special point $F=2N$ the two descriptions coincide as required for consistency. 
The dual theory allows one to completely 
understand the infrared dynamics of the very complicated strongly coupled
electric theory for $N+1<F<2N$. We also discuss the model without Yukawa
couplings which does have a conformal phase. 

Finally, we present Duality Diagrams for theories with tensors and with $SO$
and $SP$ gauge groups.

\section{SQCD with Duality Diagrams}
We introduce our notation using the ``classic" example of $N=1$ SQCD
and its dual discovered by Seiberg ~\cite{sqcd2}. In SQCD, the electric theory,
is an $SU(N)$ gauge theory with $F$ flavors of ``quarks" in the fundamental
and antifundamental representations of the gauge group. Thus the global
symmetries of the theory include two factors of $SU(F)$ acting on the
quarks and antiquarks. The tree level superpotential is taken to vanish. 
Frequently, the field content and symmetries of supersymmetric theories are
displayed using tables of Young tableaux. For example, the table for
SQCD is 

\begin{equation}
\centering
\begin{tabular}{|c||c||c|c|}\hline
field & \SU{N} & \SU{F} & \SU{F} 
\\\hline \hline
$Q$ & \Yfund & \Yfund & ${\bf 1}\vbr$ \\  \hline
$\bar{Q}$ & $\overline{\Yfund}$ & {\bf 1} & $\Yfund\vbr$ \\
\hline
\end{tabular}
\end{equation}

In the following we will frequently use such tables  to
explain our notation and demonstrate it's convenience. 
In our diagramatical notation we denote gauge groups by boxes with
the size of the gauge group stated inside the box \fbox{$\bf N\ $}.
Non-Abelian global
``flavor" symmetries are represented by the number of flavors ${\bf F }$.
Fields are represented by lines connecting the gauge and global symmetries.
For example, the quarks of SQCD are denoted by a line connecting the
box for SU(N) with the $F$ of the flavor group. Arrows on the ends of each line
distinguish representations and their complex conjugates, fundamentals have
incoming arrows whereas anti-fundamentals are denoted with outgoing arrows.

The diagram for SQCD is then 

\begin{equation}
\centering
\epsfxsize=1.5in\epsfbox{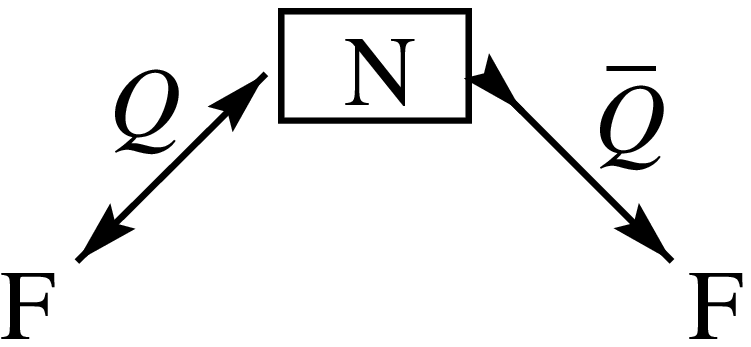}
\smallskip
\label{SQCD}
\end{equation}
where the quark and antiquark fields are labeled $Q$ and $\overline{Q}$,
respectively. From the direction of the arrows one sees that $Q$ transforms
as a fundamental under both the $SU(N)$ gauge group and the $SU(F)$ global
symmetry group. The $\overline{Q}$ is an antifundamental under $SU(N)$
and a fundamental under $SU(F)$. 

According to Seiberg's conjecture, when $F \geq N+2$, 
this theory is dual to an $SU(F-N)$ gauge
theory with $F$ flavors of dual quarks $q$ and $\overline{q}$ and a gauge
singlet field $M$ which corresponds
to the bound state $Q\overline{Q}$ of the original theory.
The ``meson'' $M$ is coupled via a tree level superpotential $W=qM\overline{q}$.
In terms of Young tableaux, this is
\begin{equation}
\centering
\begin{tabular}{|c||c||c|c|}\hline
field & \SU{F-N} & \SU{F} & \SU{F} 
\\\hline \hline
$q$ & \Yfund & $\overline{\Yfund}$ & ${\bf 1}\vbr$ \\
$\bar{q}$ & $\overline{\Yfund}$ & {\bf 1} & $\overline{\Yfund}\vbr$ \\
$M$ & ${\bf 1}$& \Yfund & $\Yfund\vbr$ \\ \hline
\end{tabular}
\label{DualSQCD}
\end{equation}

The diagram corresponding to this table is easily constructed. There is
a box for the gauge symmetry and two ${\bf F}$'s for the flavor symmetries.
Then there are two lines connecting the gauge and flavor symmetries
corresponding to $q$ and $\overline{q}$. Finally, there is the meson
line, $M$, connecting the two flavor groups. The diagram is

\begin{equation}
\centering
\epsfxsize=1.5in\epsfbox{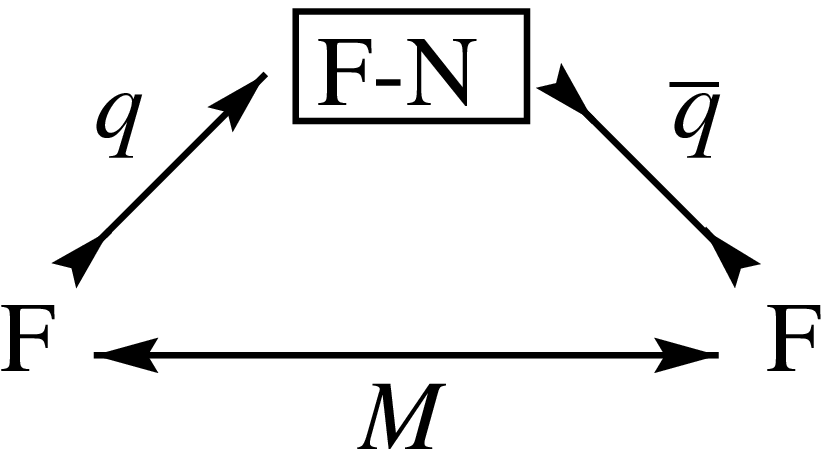}
\smallskip
\end{equation}

Note the orientation of the arrows: the arrows corresponding to the 
global groups' representations on
the two quark and antiquark lines are now reversed, the arrows for the gauge
symmetry remain the same. The arrows for the representation of 
the meson field under the global groups are pointing
in the direction in which those arrows for the fields $Q$ and $\overline{Q}$
were pointing in the original model.

We can now identify a set of ``Duality Rules" which enable us to write down
the dual diagram starting from the diagram for the electric theory.
First, the global symmetries of the model are left unchanged whilst
the gauge symmetry being ``dualized" is replaced with an as yet unknown 
dual gauge group. 
The lines corresponding to the dual quarks $q$ and $\overline{q}$ are
obtained by copying the lines corresponding to the original quarks $Q$
and $\overline{Q}$ but reversing the arrows for the global groups.
There is a meson field M in the dual 
for every pair of in- and out-going arrows of the gauge group. Here there
is one incoming arrow from $Q$ and one outgoing from $\overline{Q}$. The meson line
directly connects the two groups on which the $Q$ and $\overline{Q}$ lines ended.
The orientation of the arrows on $M$ are identical to the orientation of
the arrows on the quark lines in the orginal model. This rule is easy to understand by remembering that
$M$ is the operator map of the composite gauge invariant $Q\overline{Q}$.

Finally, the size of the dual gauge group is obtained by matching the anomalies
on the global groups. Before dualizing there were $N$ fundamentals for
each of the global groups, now there are $F$ fundamentals corresponding
to the meson $M$. Thus we need $F-N$ antifundamentals
from the dual quarks. The anomaly matching condition can be summarized
as a simple rule: the number of arrows entering any
group minus the number of arrows leaving it
has to be preserved under the duality transformation. The notation is a simple
book keeping device for the non-abelian symmetries' anomalies but the 
$U(1)$ symmetries are less transparent, one might want to keep track of
them by indicating the charges on the lines in the diagram. However,
the orginal analysis
of Seiberg \cite{sqcd2}  guarantees that the $U(1)$ anomalies
match between the electric and dual theories.

It is also possible to identify a rule that determines the superpotential
of the dual theory from the diagram. Whenever  a closed cycle
of lines,  with the arrows on the lines oriented in such a way
that all the indices can be contracted (i.e. one arrow
has to be going into a group while the other one going out), is generated in a diagram, then
the symmetries allow a superpotential term to be written. The contraction
of indices that corresponds to the flow of arrows is that using $\delta^i_j$ tensors.
In the absence of the superpotential term, the global symmetries of the
dual theory would be much larger. In particular, the global symmetries
of the meson field would not have to be the same as the dual quarks'.
The superpotential term is required to break
unwanted global symmetries whenever
there is a closed cycle of lines. 

A new situation arises when we ``dualize" our last dual again. We get an $SU(N)$
theory with new quarks and a meson $N=q\bar q$
and a term in the superpotential coupling the new quarks to the meson. The
superpotential term inherited from the first dual maps onto a mass term
for $M$ and $N$. The complete superpotential is then $W=N(M-\twi{q}\twi{q})$.
After integrating out the massive fields with their equations of motion
we recover the original theory of SQCD with vanishing superpotential.
For details see \cite{sqcd2}.

In our diagramatic notation all this is very simple. 
We ``dualize" the $SU(F-N)$ gauge group by inverting the outside arrows on the quark
lines and draw the new meson field $N$ with the arrows given by the direction of the
arrows of the quarks $q$ and $\bar q$. 

\begin{equation}
\centering
\epsfxsize=1.5in\epsfbox{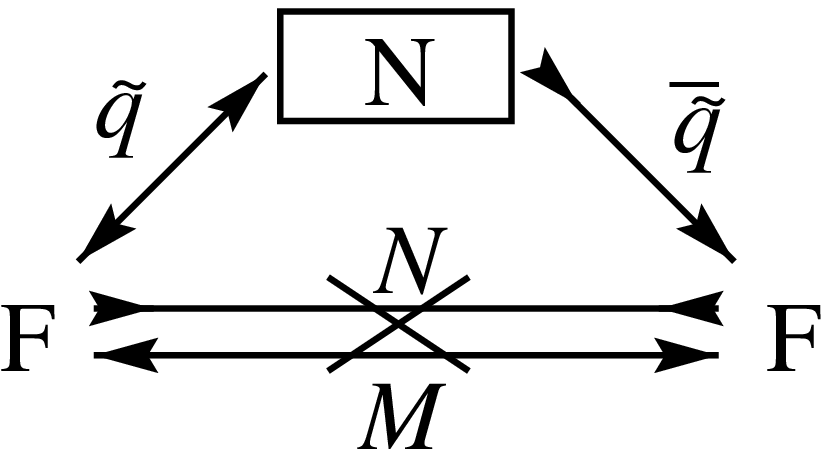}
\smallskip
\end{equation}

Note that the lines corresponding to the mesons $M$ and $N$ are parallel with opposite
arrows. This indicates that the symmetries allow (and require) a mass term for these two
fields. Thus, these fields are not present in the infrared
and should be integrated out.  We denote the mass term by crossing out
the two lines. Note that after
removing the massive fields we recover the original diagram (\ref{SQCD}) with
no superpotential
because there is no closed cycle in the diagram. In the following section
we study duals with more complicated superpotentials involving non-renormalizable
terms as well. We will see that the notation keeps track of these as well. 

Let us summarize the ``Duality Rules" in a few steps:
\begin{itemize}
\item the global symmetries of the model are unchanged, whilst the gauge group
is replaced by a dual gauge group
\item ``dualize" the quark lines by reversing the arrows on the global groups
\item draw a meson line corresponding to  every pair of in- and out-going lines
through the gauge group. The arrows on the meson line indicating its 
representations under the global groups are identical to the arrows
on the orginal quark lines
\item determine the size of the dual gauge group by matching the anomalies of
the global groups
\item each closed cycle of lines with matching arrows corresponds to a superpotential term
\item parallel lines with opposite arrows correspond to massive fields and should be
``integrated out" by simply dropping them
\end{itemize}

\subsection{Special Cases: $F \le N+1$ }

As discussed above, when $F$ falls below
$N+2$ the theory does not have a dual. Seiberg has again
provided the correct low energy description
of these theories\cite{sqcd1}. Duality Diagrams may also be used for these cases
though the notation becomes more opaque with decreasing $F$. 
For $F = N +1$, the correct low energy theory  contains the usual mesons
and fundamental baryon and anti-baryon superfields
that transform under the global symmetry groups as the dual quarks did 
for larger $F$. The Duality Rules apply
as before except that the dual gauge group is replaced by a global $U(1)$.
The superpotential term $bM \overline{b}$ 
is generated, and hence the meson and baryon fields form a triangle on the 
Duality Diagram. In addition, there is also the ``instanton term" $det(M)$
which is not encoded in the diagram in any apparent manner, and has to be
remembered separately.

\begin{equation}
\centering
\epsfxsize=3.5in\epsfbox{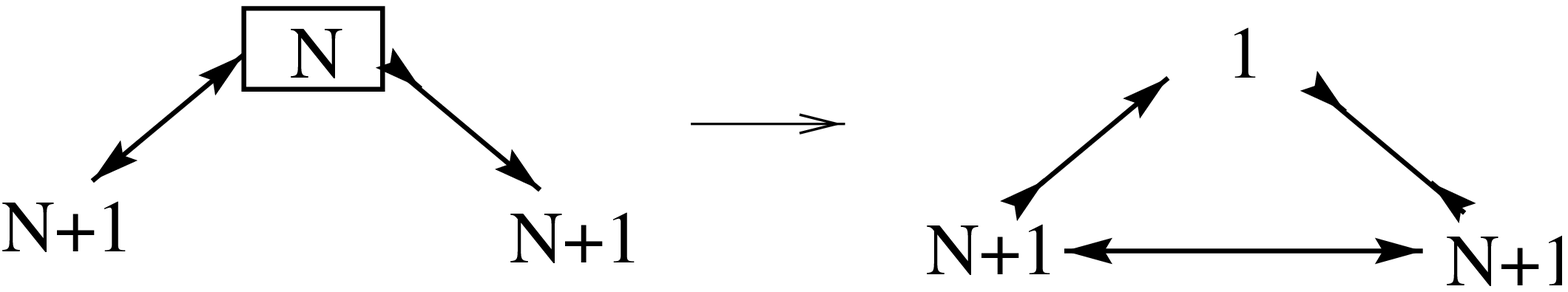}
\end{equation}

For $F=N$, there are again mesons transforming  under
the global groups but the baryon fields are now singlets under all non-abelian
groups. There is a quantum constraint
${\rm det}M - B\overline{B}=\Lambda^{2F}$ on the moduli space.
We indicate the baryons and their
constraint by drawing them as two dots over the corresponding meson line
\begin{equation}
\centering
\epsfxsize=3.5in\epsfbox{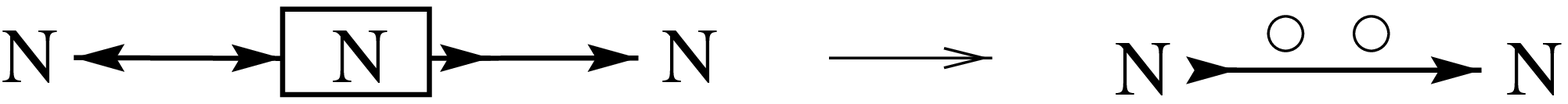}
\end{equation}

For $F < N$, the theory has no stable vacuum\cite{ADS}, thus there is no dual diagram
to draw. For $F=0$, the pure glue theory confines with no massless
degrees of freedom.
In the diagram one just erases the gauge group.

\subsection{Example: $SU(N)\times SU(M)$}
Armed with this set of rules let us now work through a non-trivial
example first studied in \cite{pop}. 
It is a theory with two $SU$ gauge groups,
various fundamentals and no superpotential.
We first define the theory with the use of our diagrams
and then, for comparison, also give the corresponding table of Young tableaux.
\beq
\centering
\epsfxsize=2in\epsfbox{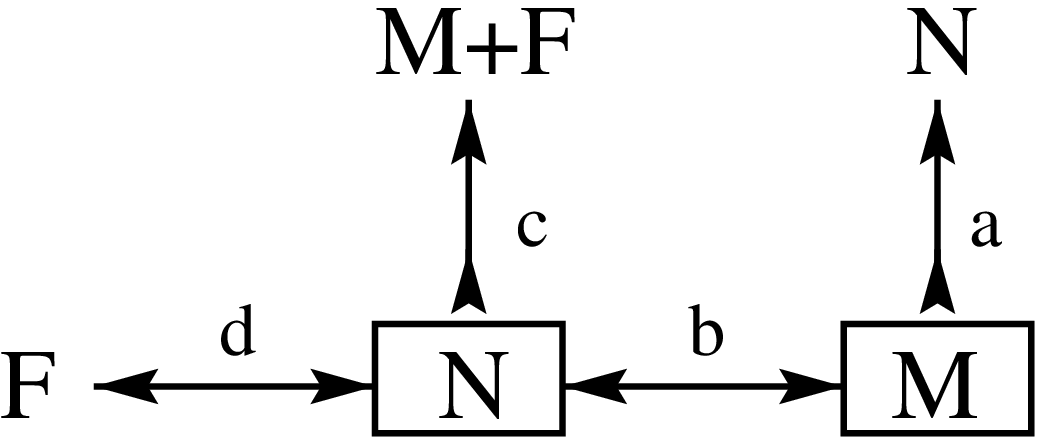}
\smallskip
\eeq

In the table
the first two groups are the gauge symmetries, the other three
are global

\begin{equation}
\label{SpSO}
\centering
\begin{tabular}{|c||c|c||c|c|c|}\hline
 Field & \SU{N} & \SU{M} & \SU{F} &\SU{M+F}& $\SU{N}$ \\
\hline\hline
  a & {\bf 1} & $\overline{\Yfund}$ & {\bf 1} & {\bf 1} & $\Yfund\vbr$\\ \hline
  b & \Yfund & \Yfund & $ {\bf 1}$ & $ {\bf 1}$ & $ {\bf 1}\vbr$ \\\hline
  c & $\overline{\Yfund}$ & {\bf 1} & {\bf 1} & \Yfund & $ {\bf 1}\vbr$ \\ \hline
  d & \Yfund & $ {\bf 1}$ & \Yfund & $ {\bf 1}$ & $ {\bf 1}\vbr$ \\ \hline
\end{tabular}
\end{equation}

We will construct duals to this theory by dualizing the two gauge groups
one at a time while treating the other as a global ``spectator" of the
duality transformation.
The justification for doing so relies on holomorphy of the infrared superpotential, $W$,
as a function of parameters of the theory.  The supersymmetric vacua
are determined by solving $V=|dW/d\phi|^2 = 0$. If such a vacuum exists
in a patch of parameter space, then a corresponding vacuum exists for
all values of parameters (except possibly some discrete points,
such as $\Lambda=0$). This follows because for a vacuum to be lifted at some
critical parameter value 
would require non-holomorphic behavior as $V$ would have to jump 
from zero. 
Therefore, there are no phase transitions, and the moduli space of vacua is
independent of the parameters of the theory. 
This allows one to take the scale of one of the gauge groups to be very
large compared to all the other scales. Then all other
gauge interactions can be treated as small perturbations. Thus one can
give a dual description of this theory by just dualizing the gauge group
with the large scale, treating the others as spectators. It then follows
from our holomorphy argument that the duality must also hold when the
scale of the dualized gauge group is not large compared to other scales
in the problem. The global anomalies of the duals obtained in this way are
guaranteed to match because the global groups are
subgroups of the global symmetry group of SQCD.
For explicit consistency checks on dualities of product group theories see \cite{pop}.

The conjecture which we want to verify is that when one alternatingly dualizes
the gauge groups of our example five times, one obtains the original
group back \cite{pop}. We impose the constraint $M-1 < N < F+M+1$
because the duals which we will encounter along the way only exist for
this range of $F$, $M$, and $N$.
These duality transformations can be performed using tables of
Young tableaux, but the tables involved
are large and the process is cumbersome. Instead, we are going to
perform all the transformations
using  diagrams and the Duality Rules. The effort
is considerably less. For comparison we give the tables and superpotentials
for each of the duals in an appendix.

First, we dualize the $SU(M)$ gauge group. Following our rules, we add a
meson field corresponding to
the gauge invariant made from a quark and an antiquark of the $SU(M)$. Then we
dualize the two quark lines by reversing the outside arrows. The dual gauge
group is $SU(N-M)$ with the following diagram
\beq
\centering
\epsfxsize=2.5in\epsfbox{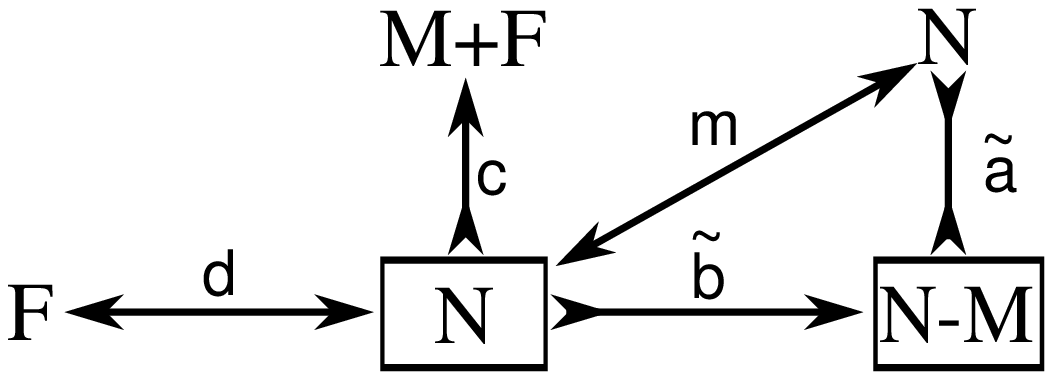}
\smallskip
\eeq

Note that there is a closed cycle with a corresponding superpotential term
$ W = m \twi{a} \twi{b} $.
Dualizing the $SU(N)$ we arrive at 
\beq
\centering
\epsfxsize=2.5in\epsfbox{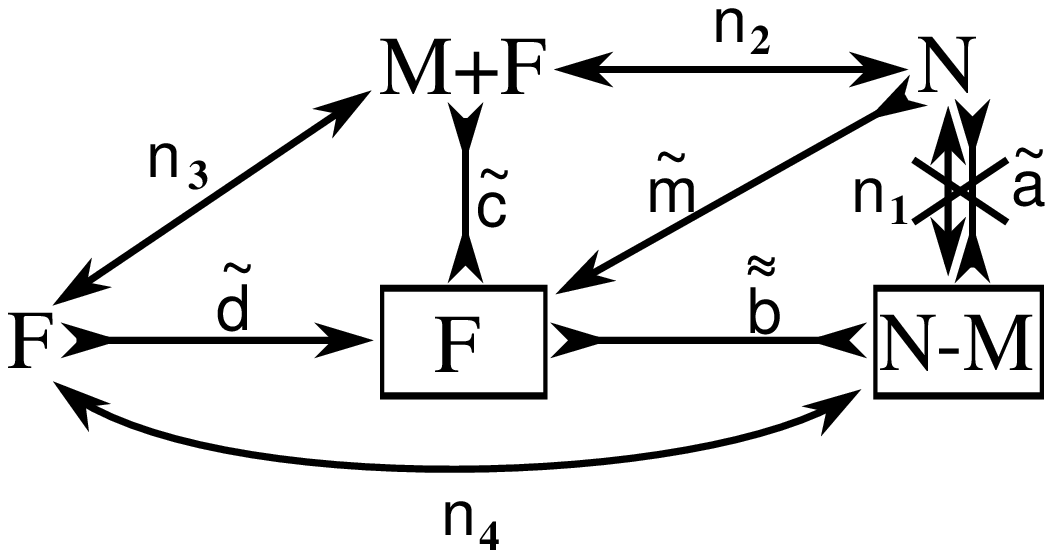}
\smallskip
\eeq

In this step four meson fields had to be added corresponding to the four
possible ways of pairing up the two ingoing and two outgoing arrows of
the $SU(N)$ gauge group. This created four new closed cycles with four
corresponding superpotential terms. 

\beq
W  =  n_1 \twi{a} + n_1 \dtwi{b} \twi{m} + n_2 \twi{c} \twi{m}
      + n_3 \twi{c} \twi{d} + n_4 \twi{d} \dtwi{b}.
\eeq

Note that the Yukawa coupling that was
present from the previous step has been turned into a mass term
because two of the fields were replaced by a composite meson. 
Integrating out the massive fields yields the theory with $n_1$ and
$\twi{a}$ dropped and replaced by their equations of
motion in the superpotential.
In the following steps, we will not explicitly write down superpotentials since
the diagramatic notation automatically keeps track of them. If in doubt,
the reader can confirm this from the appendix where we give the complete
superpotentials.

We now dualize the $SU(N-M)$ and obtain
\begin{equation}
\centering
\epsfxsize=2.5in\epsfbox{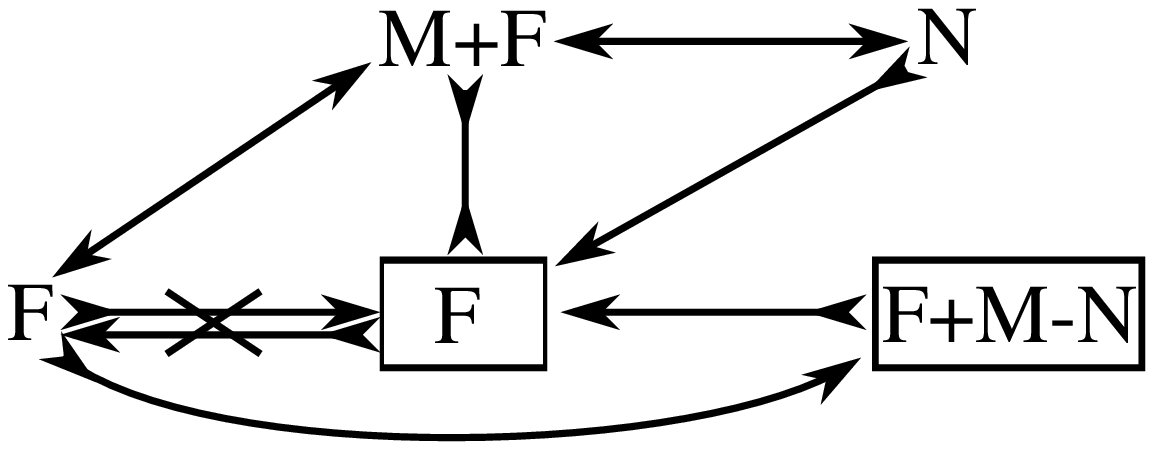}
\smallskip
\end{equation}
Note that integrating out the massive fields in this step generates a four-cycle in
the diagram which corresponds to a quartic term in the superpotential.
Dualizing $SU(F)$ we find
\begin{equation}
\centering
\epsfxsize=2.5in\epsfbox{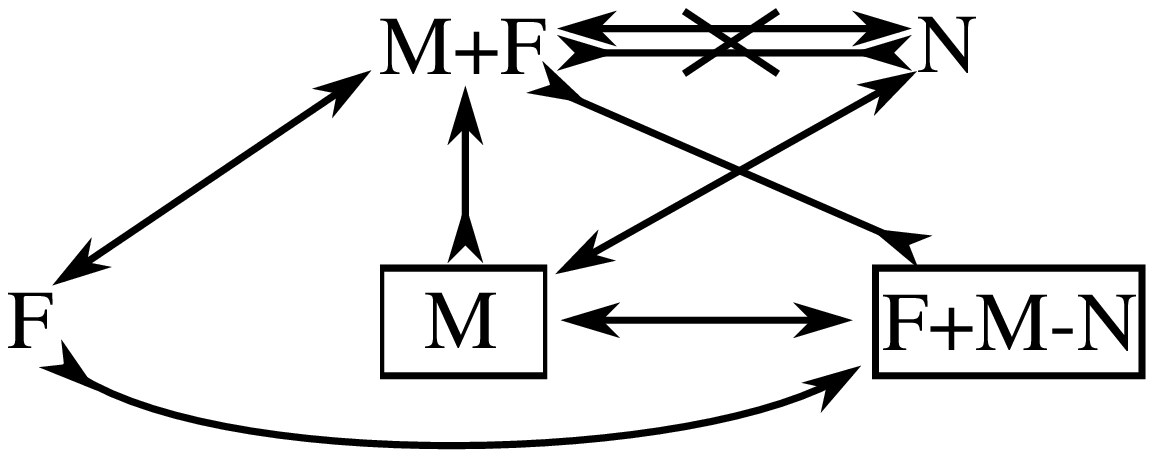}
\smallskip
\end{equation}
And finally dualizing $SU(F+M-N)$
\begin{equation}
\centering
\epsfxsize=2.5in\epsfbox{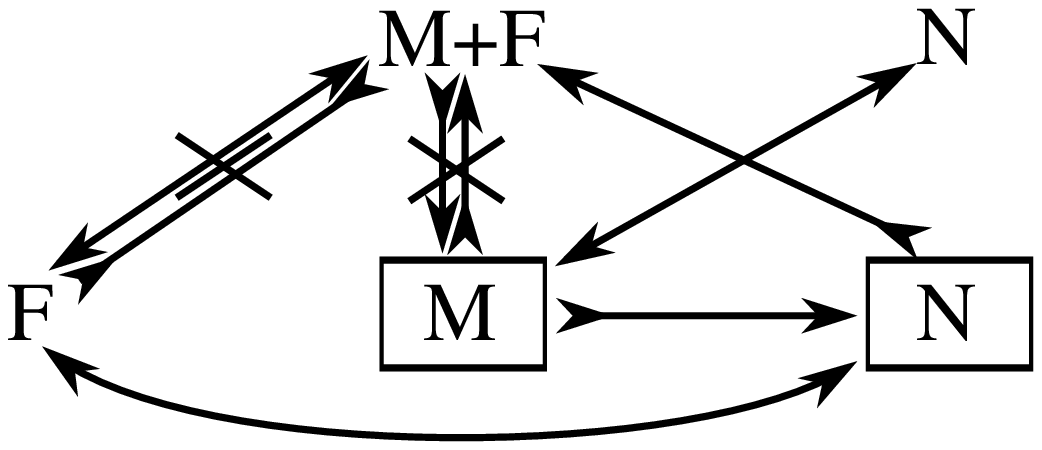}
\smallskip
\end{equation}
we return to the original theory except that the $SU(M)$ gauge group has been
replaced by its complex conjugate, and the two gauge groups changed
places.
Integrating out all the massive fields
in the last duality step has set the superpotential to zero, as required.

\section{Loop Models}

To demonstrate the power of the Duality Diagrams we next turn to the more 
complicated examples of diagrams with ``loops" of matter fields and gauge groups
which have not been studied before. The models we consider have 
$k$ $SU(N)$ gauge groups, $k$ $SU(M)$ global groups and $3 k$ chiral
superfields. These models turn out to be of interest in their own right 
since they do not possess a conformal phase. This is in contrast to the na\"{\i}ve expectation
that as long as one chooses flavors and colors such that all gauge groups taken individually
would be at a perturbative fixed point, then the whole theory should be at a perturbative
(conformal) fixed point. In our model this kind of reasoning fails dramatically.

\subsection{The Triangle, $k=3$}
The simplest non-trivial example has $k=3$. The model is then
\begin{equation} \label{loop3}
\centering
\epsfxsize=2.in\epsfbox{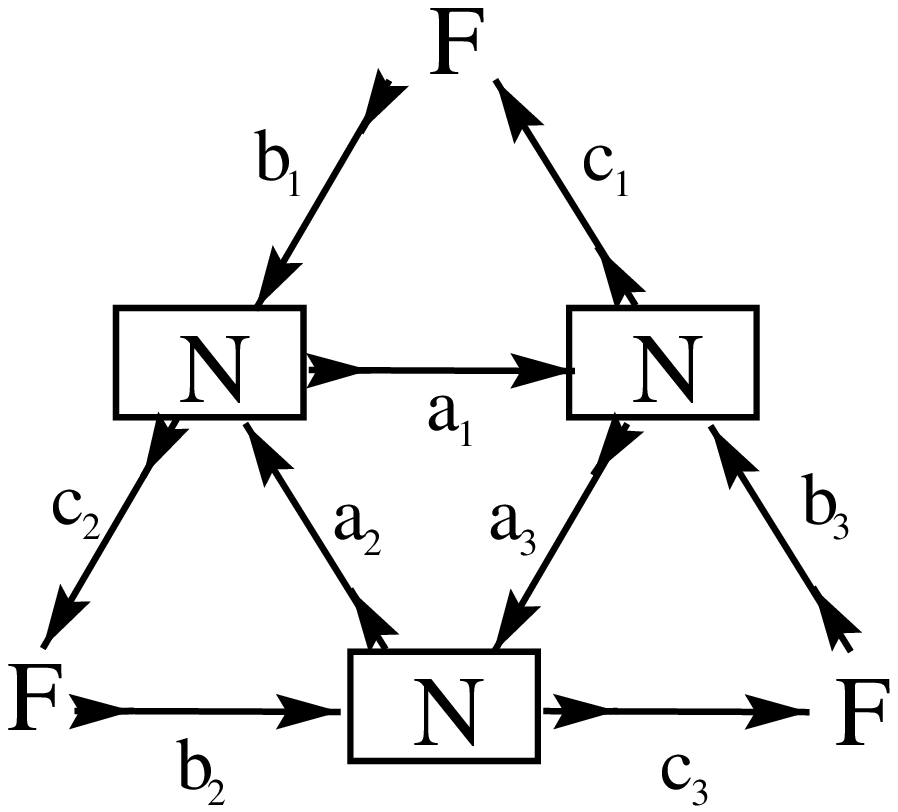}
\smallskip
\end{equation}
We  include all four Yukawa couplings compatible with the symmetries (one
associated with each triangle in the diagram).
Dualizing one of the gauge groups leads to
\begin{equation}
\centering
\epsfxsize=2.in\epsfbox{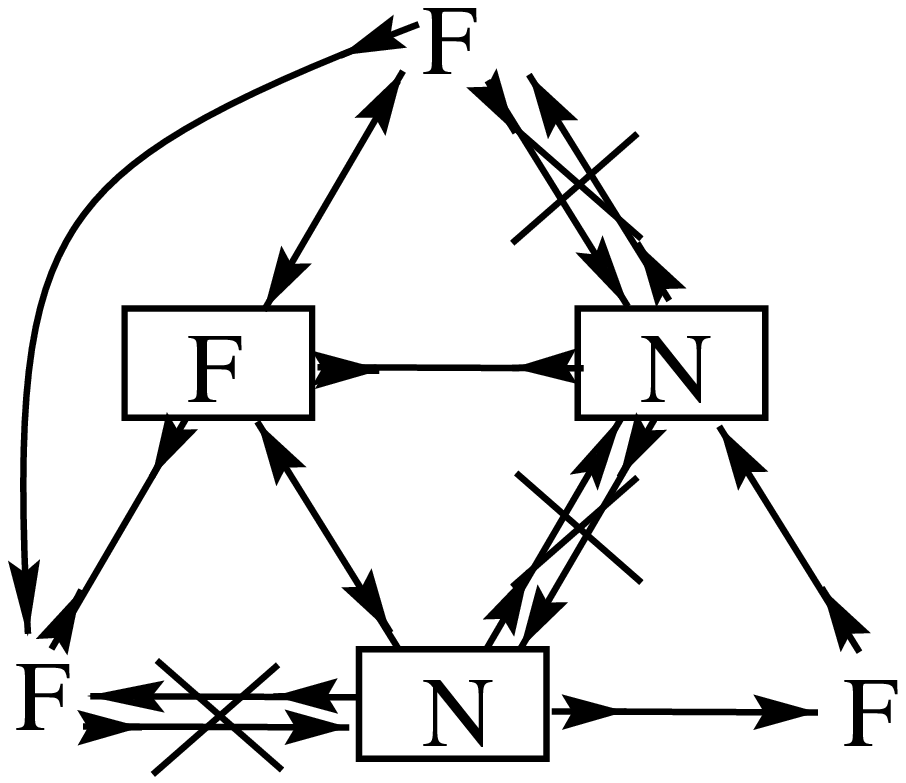}
\smallskip
\end{equation}
Next, dualizing the two remaining $SU(N)$ gauge groups in one step, we find
\begin{equation}
\centering
\epsfxsize=2.in\epsfbox{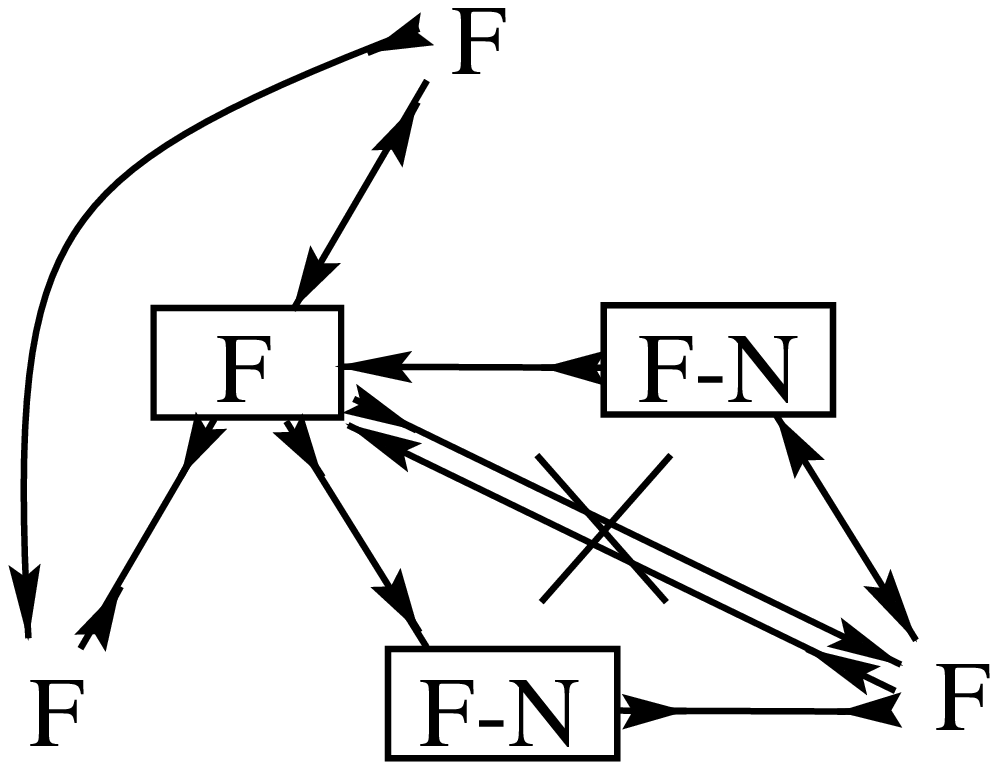}
\smallskip
\end{equation}
Finally, dualizing the $SU(F)$ gauge group gives
\begin{equation}
\centering
\epsfxsize=2.in\epsfbox{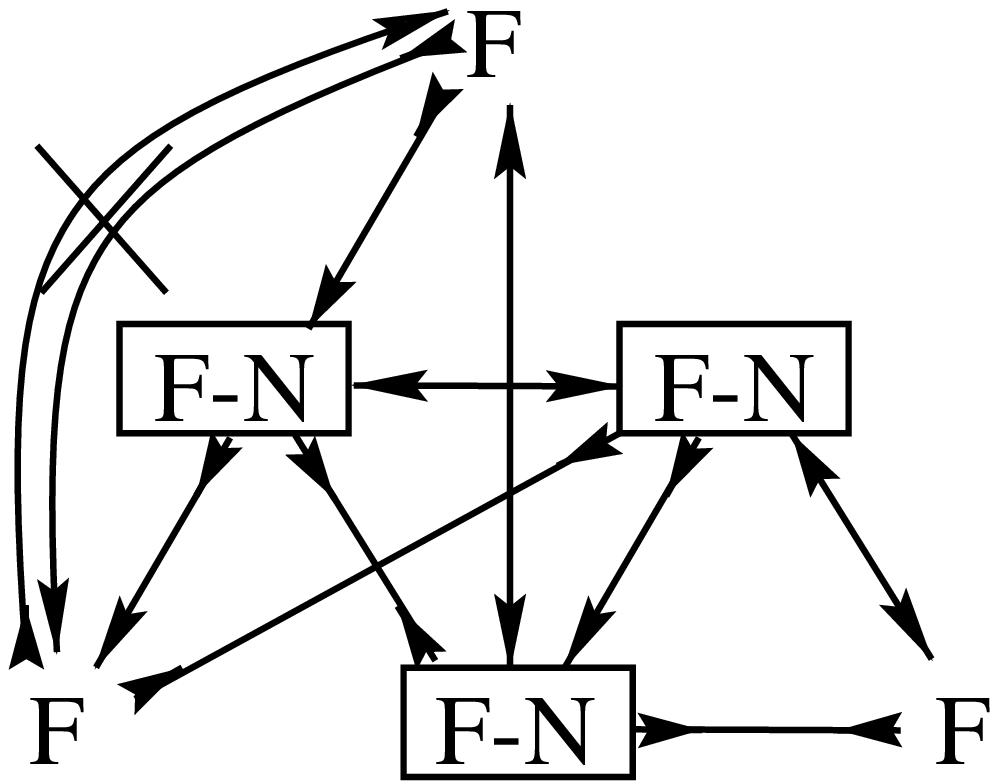}
\smallskip
\end{equation}
This dual has the same form as the orginal theory. To make this explicit
we first interchange the two gauge groups on the right of the diagram
and then complex conjugate them.
\begin{equation}
\centering
\epsfxsize=2.in\epsfbox{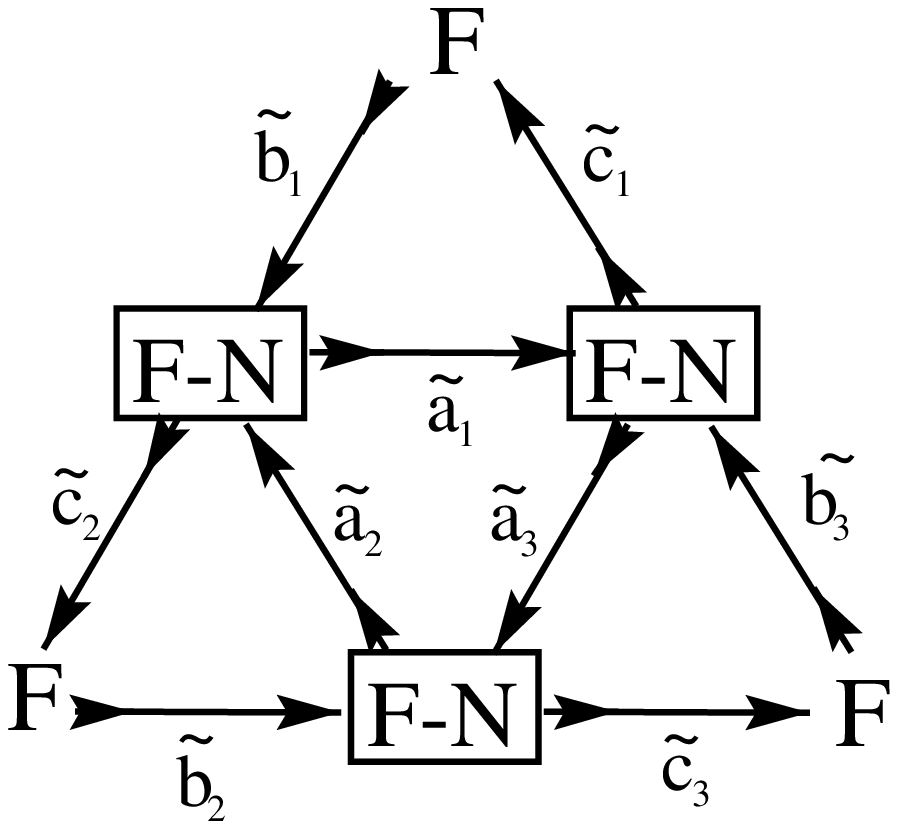}
\smallskip
\end{equation}
The dual and the electric theory are identical except that the dual
gauge group is $SU(F-N)^3$ instead of $SU(N)^3$.

Our orginal model is in a weakly coupled ``free electric" phase when it's number
of flavors ($F+N$) is greater or equal than three times the number of colors $N$,
that is when $F \ge 2N$. Similarly, the
dual magnetic theory is in it's ``free magnetic" phase when $F \le 2N$. 
At $F=2N$ it appears that we have two different
infrared-free descriptions of the same theory. Clearly, this would be a
contradiction. However, everything is consistent, because for $F = 2N$
the theory and its dual are identical.

Thus we have an infrared free 
description for any $F \geq N + 2$. This result is somewhat 
surprising since other exactly solved supersymmetric theories have
a conformal ``window"  for a certain range of flavors. For example, SQCD has a
conformal phase for $3/2 N < F < 3N$
in which neither the electric theory nor its dual are free in the infrared. 

If we did not have our dual description we might have been tempted to argue that
there should be such an interacting perturbative fixed point in the large $N$ and $F$
limit for $F$ just below $2N$. Then each of the gauge groups taken individually would be
at a perturbative Banks-Zaks fixed point\cite{bankszaks}.
However, it does not follow that the theory
as a whole is perturbative in this description as well. The problem is that all of the
chiral superfields obtain  negative anomalous dimensions through the action of the
gauge groups. Assuming that the theory is at a non-trivial fixed point
implies that the
anomalous dimension of a composite chiral operator is equal to the sum of the anomalous
dimensions of its constituents (this follows from the superconformal
R-symmetry \cite{sqcd2}), and thus the dimensions of the Yukawa  couplings are
less than three, rendering the Yukawa couplings relevant. Therefore the Yukawa
couplings are large in the infrared and a perturbative analysis like the one we attempted
is invalid. 

A single group in the model discussed here can be placed in an 
approximate conformal phase by taking its scale $\Lambda$ much greater
than those of the other two groups. However, eventually the other two groups
become strong at low energies and contribute to 
anomalous dimensions of the fields which appear in the
beta function of the original group.
This can be seen in the duals of the theory where dualizing one group
leads to a change in the number of flavors of the other groups.
The gauge and Yukawa couplings interfere so
that the model does not have a non-trivial fixed point in the infrared.

Another indication that the theory is not at a conformal fixed point, comes from the fact
that for $F \ne 2N$ the superpotential couplings break all non-anomalous R-symmetries.
Thus there is no candidate for the anomaly free superconformal R-symmetry
which is part of the superconformal algebra\footnote{This argument is
not completely rigorous because the superconformal R-symmetry might be
an accidental symmetry of the infrared.}.

Even though the construction of the dual via a series of elementary dualities
guarantees consistency at each step, it is interesting to perform some consistency
checks on the final dual. In addition to the non-abelian $SU(F)$ flavor
symmetries, the theory also possesses three non-anomalous $U(1)$ symmetries.
As a first consistency check, we find that all global anomalies match.
Furthermore, we can define a map identifying each non-vanishing gauge-invariant
operator of the original theory with a partner in the dual. For the
meson fields this operator map is trivial, mapping the mesons
$b_1 c_2, b_2 c_3, b_3 c_1$ onto $\twi{b_1} \twi{c_2},
\twi{b_2} \twi{c_3}, \twi{b_3} \twi{c_1}$, respectively.
An interesting consistency check arises when we integrate out flavors in one
of the theories. 
Let us, for example, add mass terms for one set of flavors in the electric theory.
The superpotential is then
\beq
W  =  a_1 a_2 a_3 + \sum_{i=1}^3 a_i b_i c_i + b_1 c_2 + b_2 c_3 + b_3 c_1,
\eeq
where the flavor indices are summed over in the $a_i b_i c_i$ terms, whereas
the mass terms are only for the $F$'th flavors. The equations of motion for
the massive fields yield two different branches of vacua. 
The ``mass branch" with vanishing vacuum expectation values for the massive
$b_i$ and $c_i$ simply reduces the number of flavors by one and leaves
the gauge groups unchanged.
The other branch, the ``Higgs branch", has expectation values
$\langle a_1 a_2 a_3\rangle =1$
and non-zero expectation values for the $b_i$ and $c_i$.
In this branch, the gauge groups are broken to $SU(N-1)$ and the number of flavors
is reduced to $F-1$.
It is interesting to see how this is realized in the dual. The mass terms simply map
onto corresponding mass terms $\twi{b_1} \twi{c_2} + \twi{b_2} \twi{c_3}
+ \twi{b_3} \twi{c_1}$. Again, there are two branches. But the correspondence
between the branches in the two descriptions is non-trivial: the ``mass branch"
of the electric theory is mapped onto the ``Higgs branch" of the magnetic theory
and vice versa. This non-trivial mapping is required to preserve the duality
$N \leftrightarrow \twi{N}=F-N$ and $F \leftrightarrow \twi{F}=F$ under
the mass perturbation.

We conclude this section by briefly stating results for the theories with $F<N+2$
which are very similar to results in SQCD with the corresponding numbers
flavors\cite{sqcd1}.
For $F=N+1$, the dual has no gauge group and the theory is confining. In the
infrared the theory is described by mesons and baryons interacting via a confining
superpotential. The diagrams then don't enable one to keep track of the
determinant terms in the superpotential. This is analogous to SQCD with
$F=N+1$. 
For $F=N$, the theory confines with chiral symmetry breaking, again the discussion
is very similar to SQCD with $F=N$. Finally, for $F < N$, nonperturbatively
generated superpotentials lift the potential at the origin. The vacuum is stabilized
by our Yukawa couplings, and we find supersymmetric
vacua at non-zero expectation values for the fields. The remaining unbroken gauge
groups $SU(N-F)$ lead to gaugino condensation.

\subsection{Loops with $k \ge 3$}

The model may be extended to include $k$ $SU(N)$ gauge groups in the loop
and $k$ $SU(F)$ global groups in the same pattern as the model above. 
We again include Yukawa interactions for all the superfields transforming under
the global symmetry groups. The Yukawa interaction for the central loop of
the model in (\ref{loop3}) is now replaced by a (non-renormalizable) 
dimension $k$ operator
involving each of the fields in the loop.

We now present a simple inductive proof to show that these models also have duals 
of the same loop form but with dual gauge group $SU(N-F)$. The case $k=3$ provides
the first step for our induction. Assume that the case $k-1$ has been proven. Now, consider
a segment of the loop of the $k$'th model
\begin{equation}
\centering
\epsfxsize=3in\epsfbox{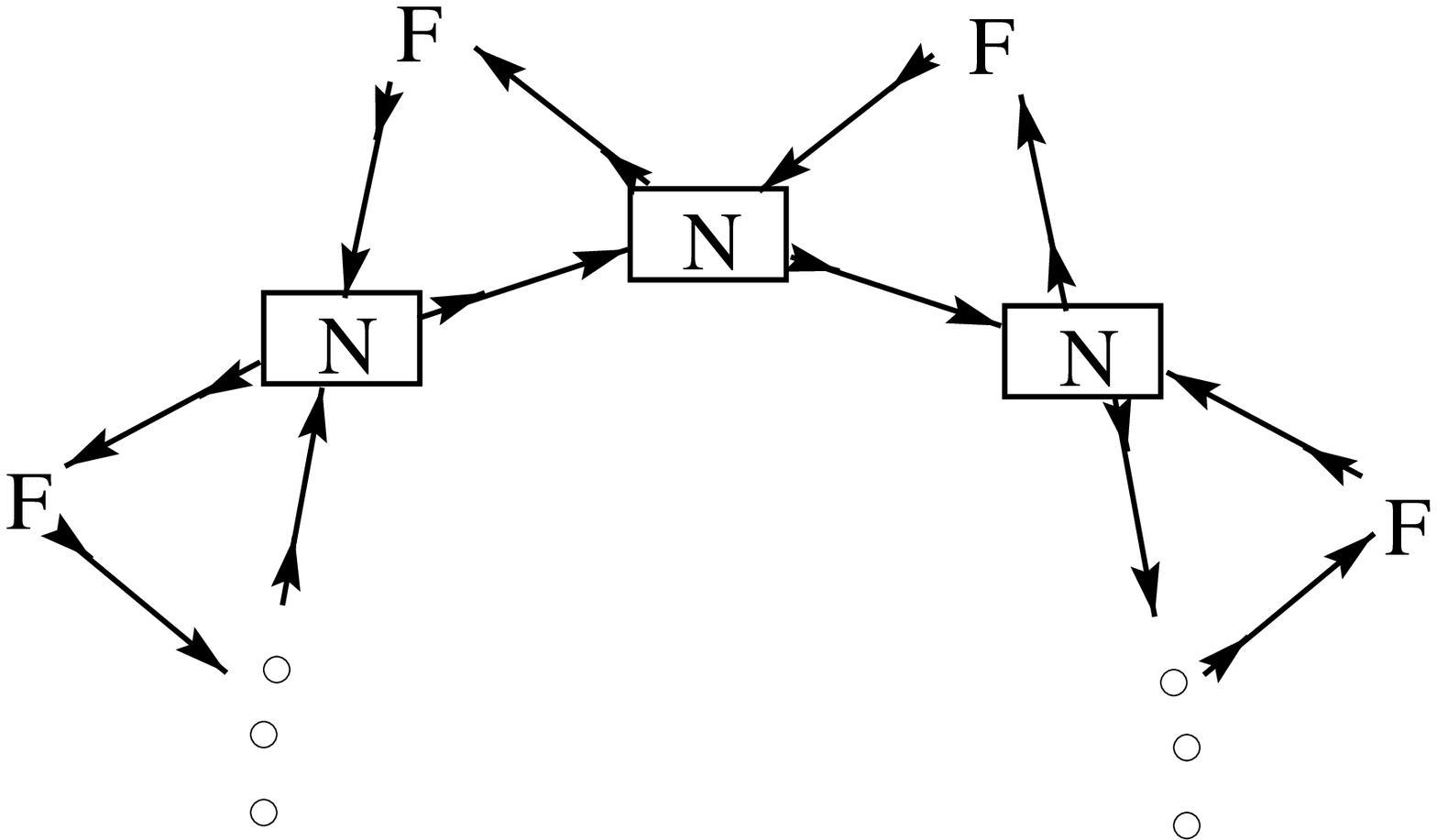}
\smallskip
\end{equation}
We begin by dualizing an $SU(N)$ gauge group
\begin{equation}
\centering
\epsfxsize=3in\epsfbox{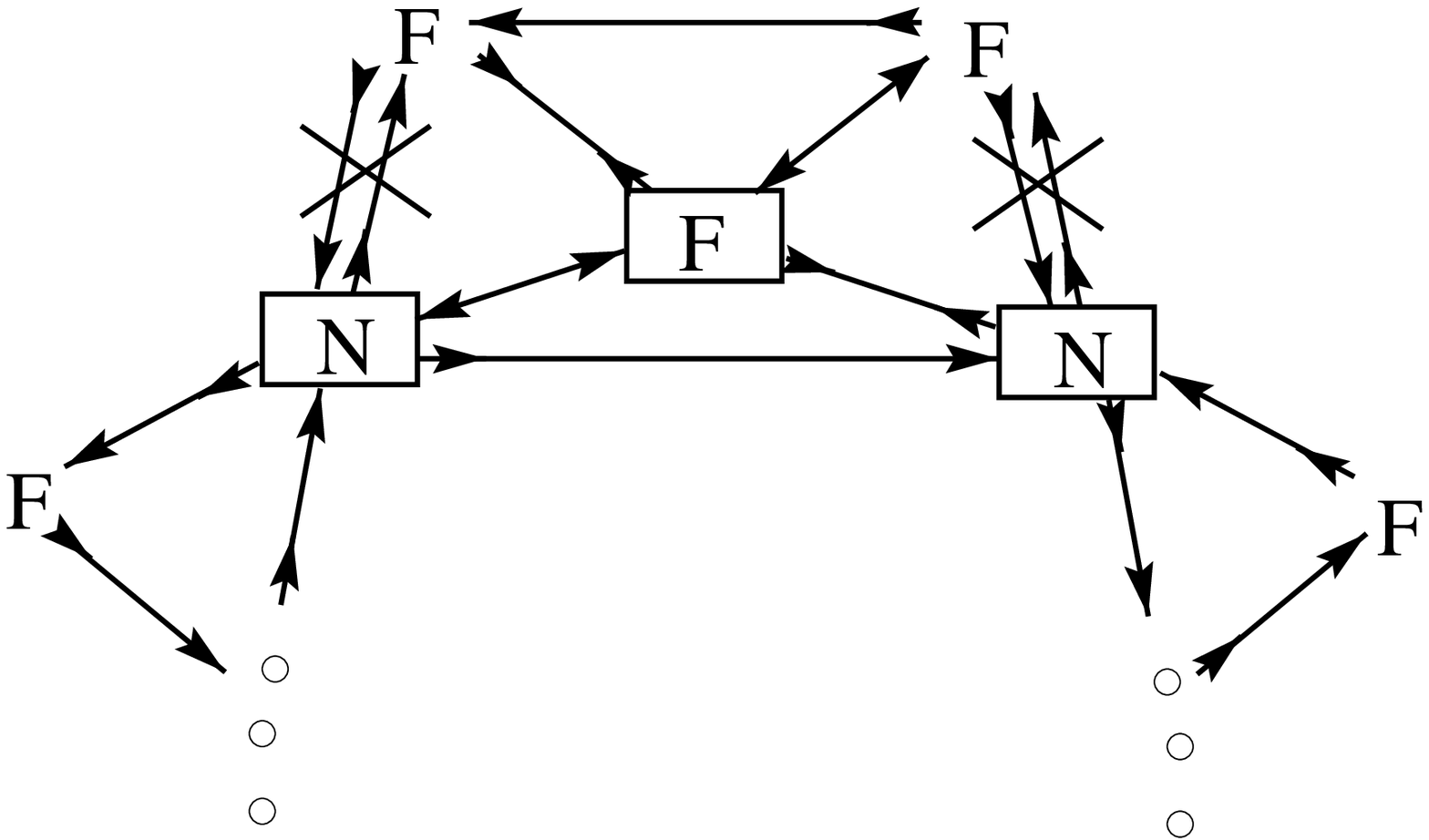}
\smallskip
\end{equation}
The resulting dual has the form of a loop model with $k-1$ $SU(N)$ gauge groups
but where one $SU(F)$ global group is the dual gauge group just produced. 
In addition, there are some superfields that do not interact with the
rest of the loop.
Then, treating the $SU(F)$ as a ``spectator", we can apply the dual for $k-1$ and obtain
\begin{equation}
\centering
\epsfxsize=3in\epsfbox{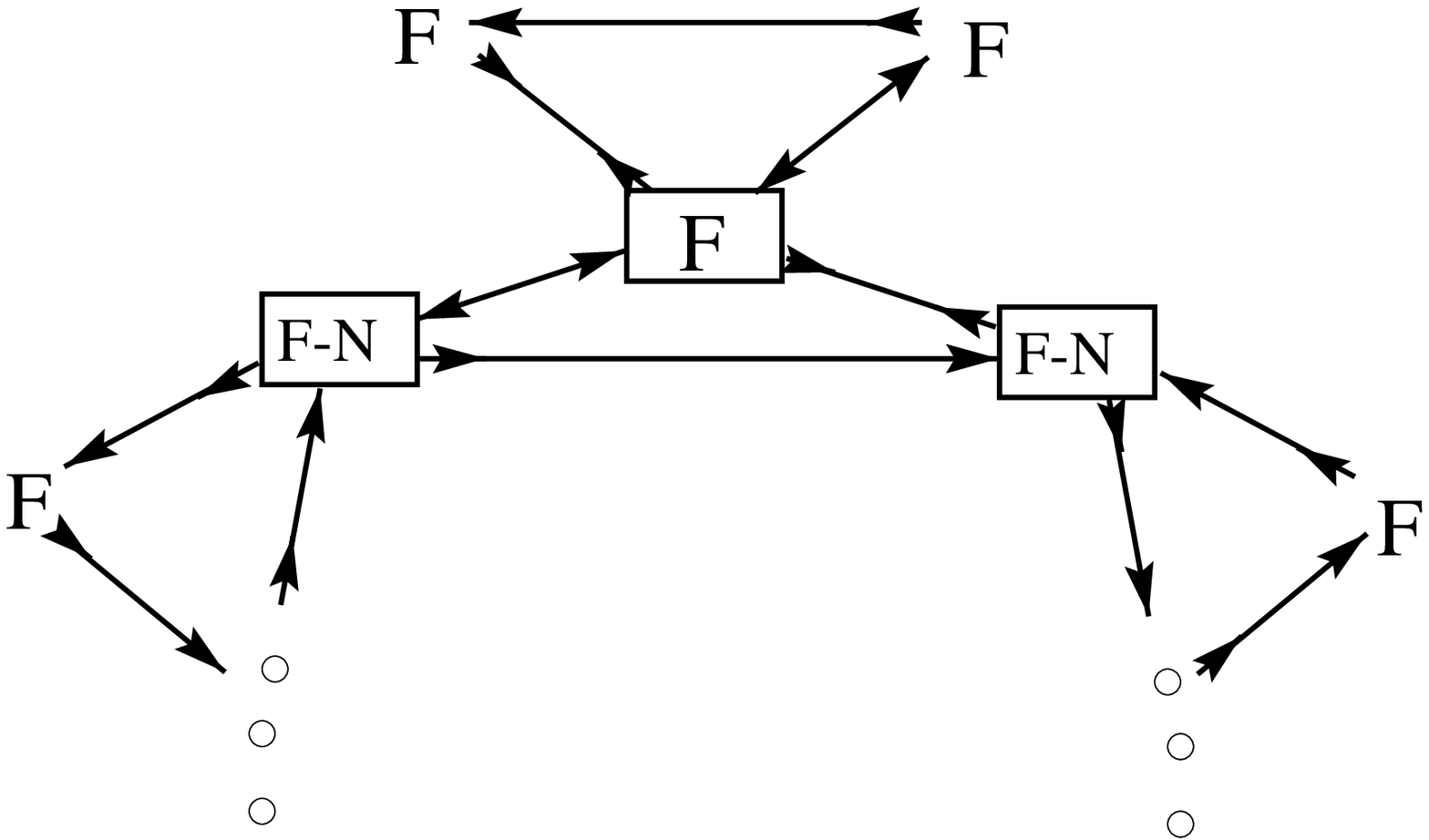}
\smallskip
\end{equation}
Now dualizing the $SU(F)$ group gives
\begin{equation}
\centering
\epsfxsize=3in\epsfbox{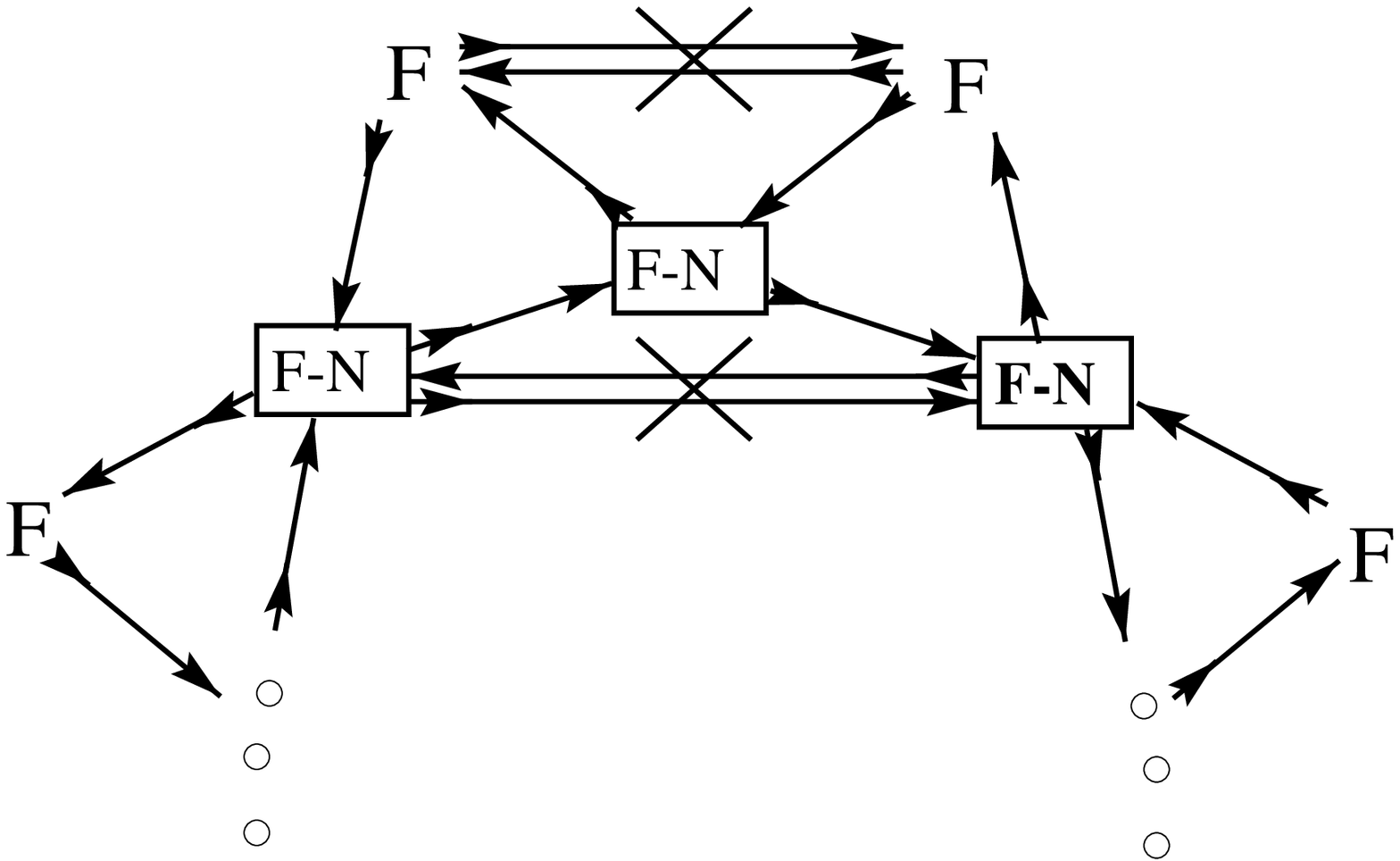}
\smallskip
\end{equation}
and we see that the $k$'th loop has a dual of the form advertised. In total, to reach this
dual description it takes $2(k-1)$ separate steps of dualizing subgroups. 
The larger $k$ loop models also have the property that the electric description
is infrared-free for $F \ge 2N$, whereas the magnetic description is free for
$F \le 2N$. They have no conformal phase.

The case $k=2$ is special, since the ``Yukawa coupling" for the central
loop is now a mass term. On integrating out the massive
multiplets the Yukawa terms of the fields transforming under the global groups 
generate a quartic interaction involving the remaining fields. The two gauge 
groups may then be dualized, and the two resulting mesons obtain masses
from the operator map of the quartic term.
The dual with gauge groups $SU(F-N)$ is similar to the
original theory again. However, the range of flavors for which the two descriptions
are free differ from the $k>2$ cases. The electric theory is free for $F \ge 3N$
whereas the magnetic theory is free for $F \le 3/2N$.
For $3/2N<F<3N$ the theory flows to an
interacting fixed point in the infrared.

\subsection{Related Models}

The models and their duals take a different form if we remove the Yukawa couplings
between the superfields transforming under the global groups. Removing these 
interactions increases the global symmetry of the model which then takes the form
(for $k=3$)
\begin{equation}
\centering
\epsfxsize=2.5in\epsfbox{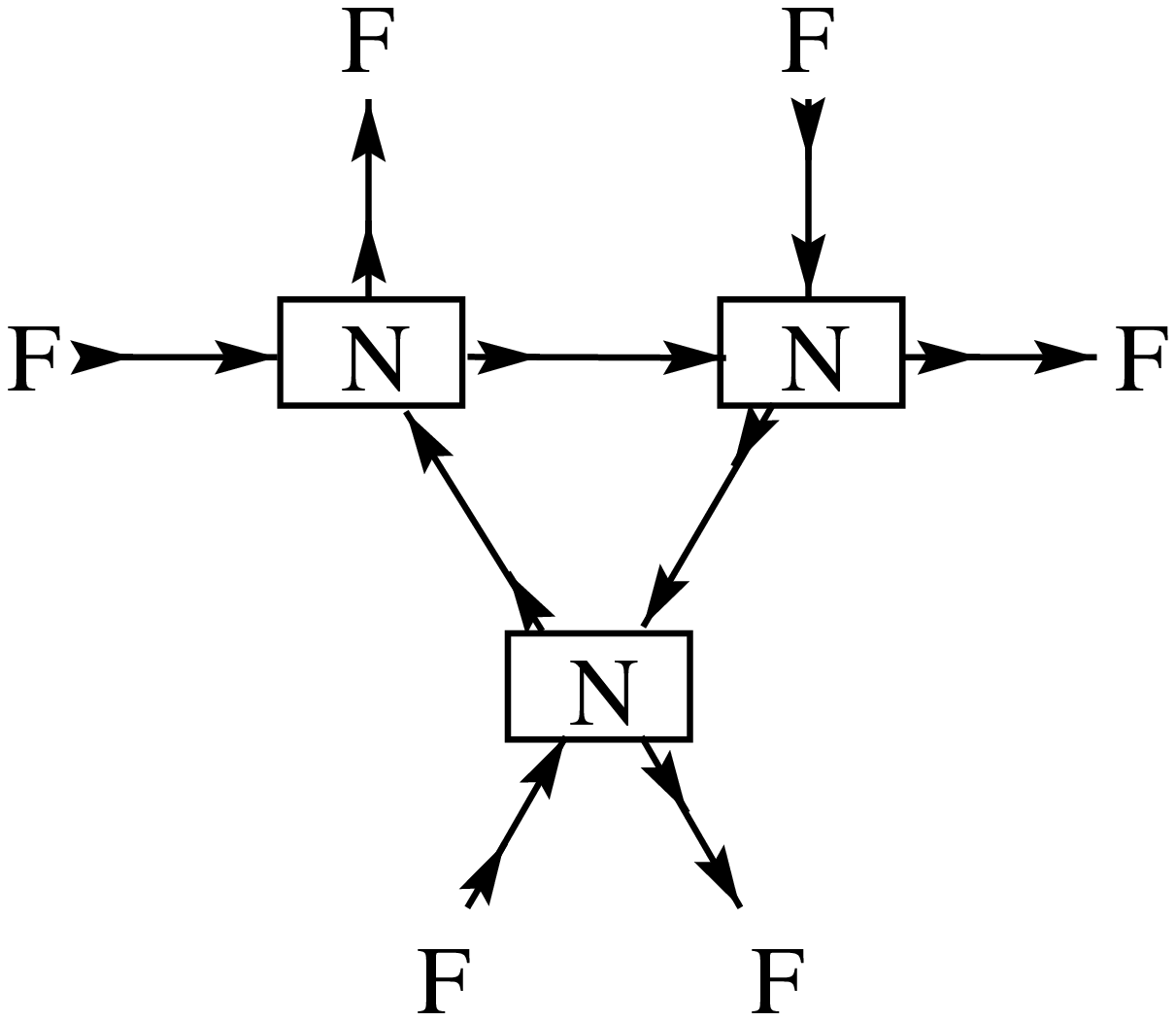}
\smallskip
\end{equation}
The model is again brought to an interesting dual by dualizing the subgroups $2(k-1)$
times.
This dual has $k$ $SU((k-1)F-N)$ gauge groups, $2k$ $SU(F)$ global groups, and 
additional mesons connecting the global groups. For $k=3$ the dual is
\begin{equation}
\centering
\epsfxsize=2.5in\epsfbox{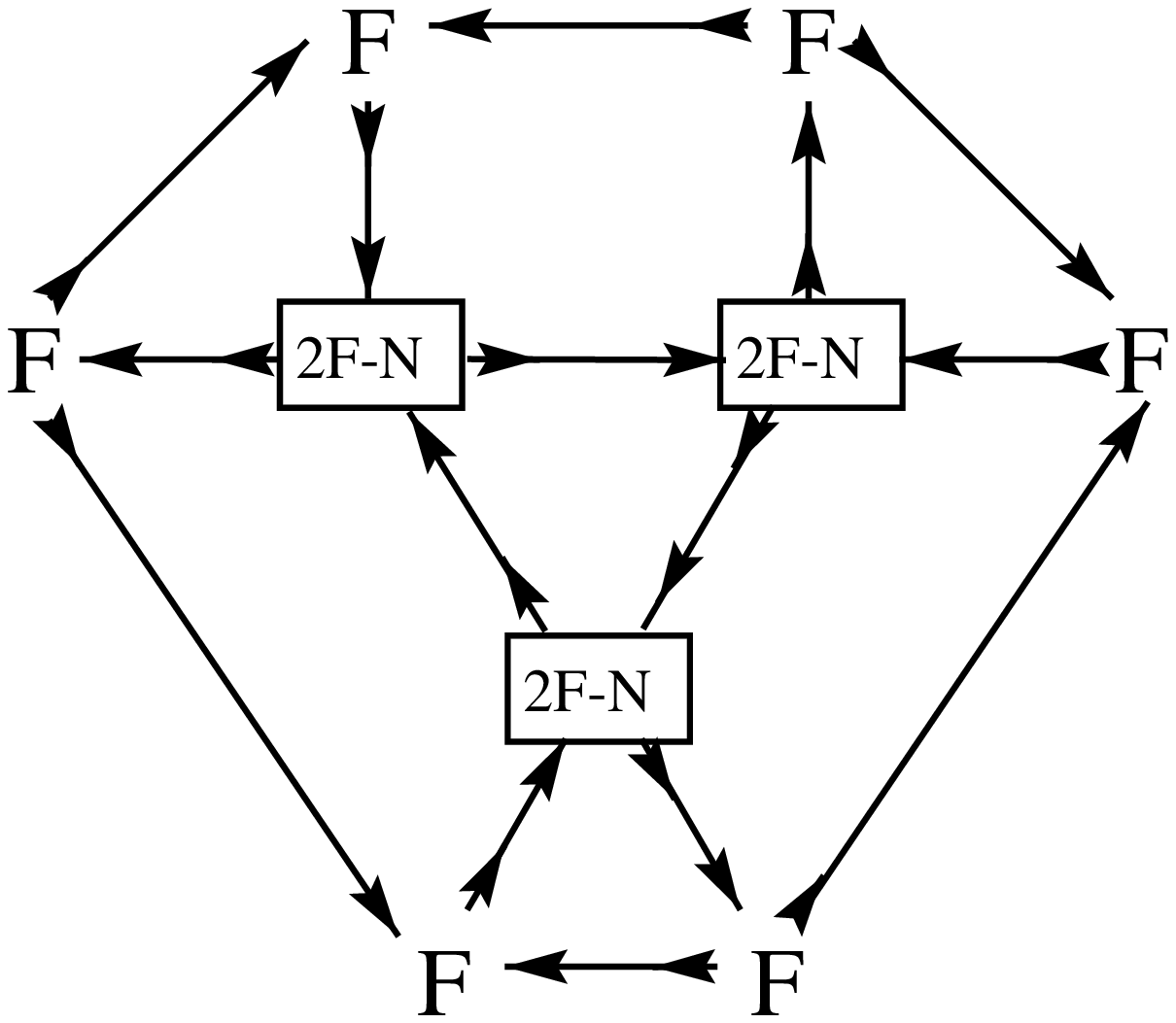}
\smallskip
\end{equation}
Unlike its sister (\ref{loop3}) this model does have a conformal 
phase for $2N/3<F<2N$. In the general case, the conformal range extends from
$F=2N$ down to $F=2N/(2k-3)$. Again, this is different from what one would have
expected from treating each of the gauge groups separately. Then one would
have falsely concluded that the conformal range is $N/2<F<2N$.

\section{Duality Diagrams for SO and Sp groups and tensors}

So far, we have presented our notation only for $SU$ groups with
matter transforming in the fundamental and antifundamental
representations. The notation is easily extended to deal with $SO$ and $Sp$
gauge groups. We simply append an index indicating the different groups.
For an $Sp(2N)$ gauge group\footnote{In our convention, $Sp(2N)$ is the
group whose fundamental representation is $2N$ dimensional.
In the literature, this group is sometimes referred
to as $Sp(N)$.} we write ${\bf \fbox{2N}_{Sp}}$, and a global $SO(F)$ symmetry
would be denoted as ${\bf F_{SO}}$. Since the representations of $Sp$ and $SO$
are (pseudo)real we do not need to put any arrows on the ends of lines
connected to these groups. For example, a field transforming as a
fundamental under an $SO(N)$ gauge group and a fundamental under a global
$SU(F)$ would be ${\bf \fbox{N}_{SO} \longrightarrow F}$. Sometimes, to avoid
confusion, it might be useful to use a subscript to denote $SU$ groups as well.

When working with these groups one often encounters fields transforming
as symmetric and antisymmetric two-index tensors.
To denote them we also use lines but we
put circles with their cubic anomaly coefficient on the lines. The cubic
anomaly is sufficiently unique to distinguish almost all tensor
representations.In those cases where it is not sufficient (for 
example $Sp$ and $SO$ gauge groups with only (pseudo)real representations
where these indices all vanish) 
the Young tableaux may be used instead to denote
the representation. This enables one to
quickly calculate the cubic anomalies associated with any of the non-abelian
groups in the diagram. For example, a field transforming as an antisymmetric two
index tensor under a gauged $SU(N+4)$ would be 
\begin{equation}
\centering
\epsfxsize=1.2in\epsfbox{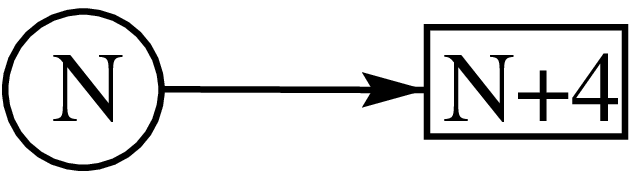}
\smallskip
\end{equation}
To retain the connection between superpotential terms and closed loops
in more complicated models is difficult. One possibility is to introduce
an extra bit of notation for higher
dimension representations. Every time a meson which is a higher
dimension representation of a global group is generated by a duality
transformation a dotted line may be included linking it to the gauge
group that confined its constituents. The dotted line now closes
a cycle indicating a superpotential term involving the meson and the
other fields in the loop, though the precise form of the superpotential
term is not explicit from the diagram. It may be necessary to keep
track of the superpotential separately.

In this notation, the duals for $SO$ and $Sp$ gauge theories with
fundamentals \cite{sospdual}, respectively, are given by
\begin{equation}
\centering
\epsfxsize=1.5in\epsfbox{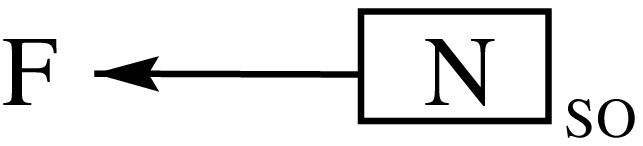}\hspace{1cm}\longrightarrow \hspace{1cm}
\epsfxsize=2in\epsfbox{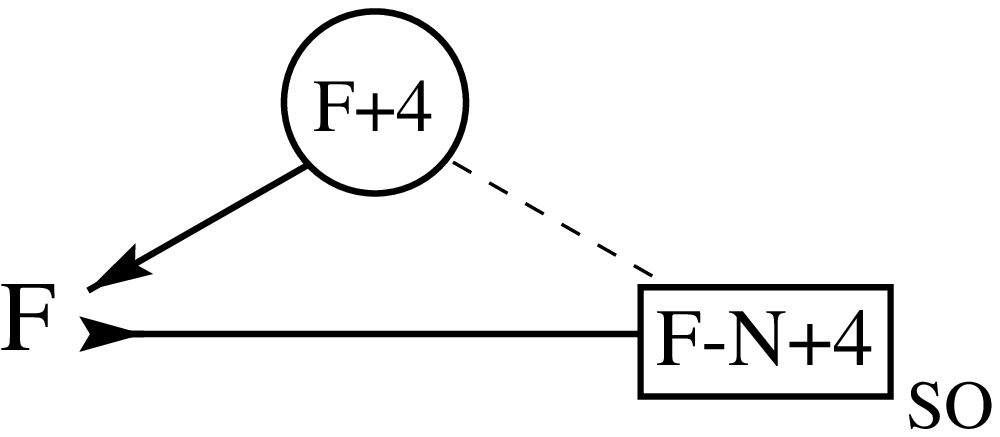}
\smallskip
\end{equation}
\begin{equation}
\centering
\epsfxsize=1.5in\epsfbox{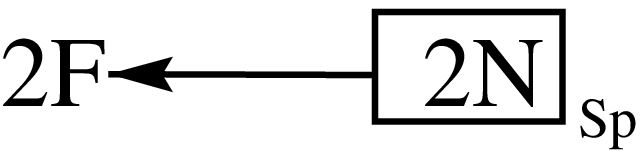} \hspace{1cm} \longrightarrow \hspace{1cm}
\epsfxsize=2.in\epsfbox{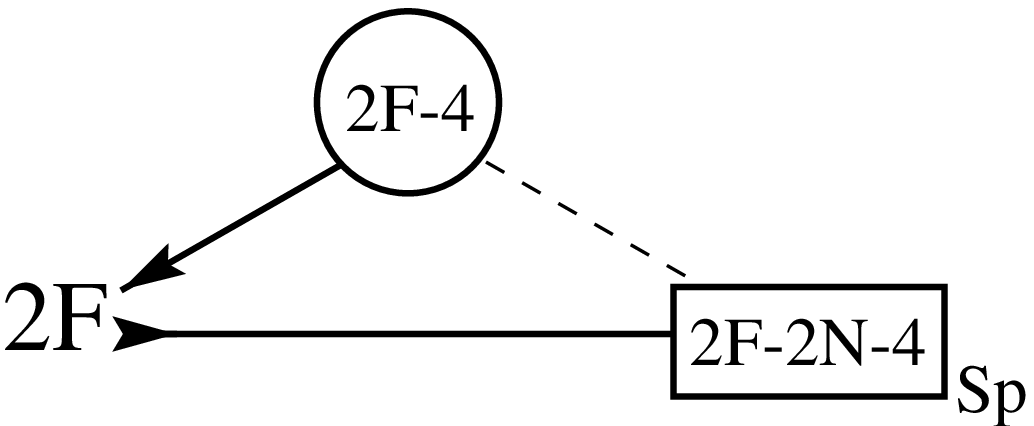}
\smallskip
\end{equation}
%

\section*{Acknowledgments}
We would like to thank A. Cohen and D. Kaplan
for helpful discussions, and C. Csaki and W. Skiba for comments on the
manuscript. We also thank the Aspen Center for Theoretical
Physics where this work was initiated.
This work was supported in part by the Department of Energy under
contracts \#DE-FG02-91ER40676, \#DE-FG02-91ERU3075 and \#DE-AC02-76ER03069 and
by the National Science Foundation grant \#PHY89-04035 .

\newpage

\appendix
\section{Tables for the ${\bf SU(N) \times SU(M)}$ Model}
In this appendix, we give tables of representations and superpotentials
for the model given in section 2.2. We denote fields that have
been ``dualized'' once by a tilde, and fields that have been dualized
$i$ times by an upper index $^{(i)}$. After each duality step we list the
full particle content and superpotentials including all fields,
followed by the superpotential obtained by integrating out the massive fields.

\begin{equation}
\centering
\begin{tabular}{|c||c|c||c|c|c|}\hline
 Field & \SU{N} & \SU{M} & \SU{F} &\SU{M+F}& $\SU{N}$ \\
\hline\hline
  a & {\bf 1} & $\overline{\Yfund}$ & {\bf 1} & {\bf 1} & $\Yfund\vbr$\\ \hline
  b & \Yfund & \Yfund & $ {\bf 1}$ & $ {\bf 1}$ & $ {\bf 1}\vbr$ \\\hline
  c & $\overline{\Yfund}$ & {\bf 1} & {\bf 1} & \Yfund & $ {\bf 1}\vbr$ \\ \hline
  d & \Yfund & $ {\bf 1}$ & \Yfund & $ {\bf 1}$ & $ {\bf 1}\vbr$ \\ \hline
\end{tabular}
\end{equation}
\begin{equation}
 W = 0 
\end{equation}\vskip-.6cm

\begin{equation}
\centering
\begin{tabular}{|c||c|c||c|c|c|}\hline
 Field & \SU{N} & \SU{N-M} & \SU{F} &\SU{M+F}& $\SU{N}$ \\
\hline\hline
  $\tilde{a}$ & {\bf 1} & $\overline{\Yfund}$ & {\bf 1} & {\bf 1} &
  $\overline{\Yfund}\vbr$\\ \hline
  $\tilde{b}$ & $\overline{\Yfund}$ & \Yfund & $ {\bf 1}$ & $ {\bf 1}$ & $ {\bf 1}\vbr$ \\\hline
  c & $\overline{\Yfund}$ & {\bf 1} & {\bf 1} & \Yfund & $ {\bf 1}\vbr$ \\ \hline
  d & \Yfund & $ {\bf 1}$ & \Yfund & $ {\bf 1}$ & $ {\bf 1}\vbr$ \\
  \hline
  m & \Yfund & $ {\bf 1}$ & {\bf 1} & $ {\bf 1}$ & $ \Yfund\vbr$ \\ \hline
\end{tabular}
\end{equation}
\begin{equation} 
W = m \tilde{a} \tilde{b} 
\end{equation}\vskip-.6cm

\begin{equation}
\centering
\begin{tabular}{|c||c|c||c|c|c|}\hline
 Field & \SU{F} & \SU{N-M} & \SU{F} &\SU{M+F}& $\SU{N}$ \\
\hline\hline
$\tilde{a}$ & {\bf 1} & $\overline{\Yfund}$ & {\bf 1} & {\bf 1} &
  $\overline{\Yfund}\vbr$\\ \hline
$b^{(2)}$ & $\overline{\Yfund}$ & $\overline{\Yfund}$ & $ {\bf 1}$
  & $ {\bf 1}$ & $ {\bf 1}\vbr$ \\\hline
$\tilde{c}$ & $\overline{\Yfund}$ & {\bf 1} & {\bf 1} &
 $ \overline{\Yfund}$ 
  & $ {\bf 1}\vbr$ \\ \hline
$\tilde{d}$ & \Yfund & $ {\bf 1}$ & $\overline{\Yfund}$ & 
$ {\bf 1}$ & $ {\bf 1}\vbr$ \\
  \hline
$\tilde{m}$ & \Yfund & $ {\bf 1}$ & {\bf 1} & $ {\bf 1}$ & 
  $\overline{\Yfund} \vbr$ \\ \hline
$n_1$ & ${\bf 1}$  & $\Yfund $ & {\bf 1} & $ {\bf 1}$ & 
  $\Yfund \vbr$ \\ \hline
$n_2$ & ${\bf 1}$ &  ${\bf 1}$ & ${\bf 1}$ & $ \Yfund$ & 
  $\Yfund \vbr$ \\ \hline
$n_3$ & ${\bf 1}$ & $ {\bf 1}$ & $\Yfund$ & $ \Yfund$ & 
 $ {\bf 1}  \vbr$ \\ \hline
$n_4$ & ${\bf 1}$ & $\Yfund$  & $\Yfund$  & $ {\bf 1}$ & 
  ${\bf 1} \vbr$ \\ \hline
\end{tabular}
\end{equation}
\beq 
W = n_1 \tilde{a} + n_1 b^{(2)} \tilde{m} + n_2 \tilde{c} \tilde{m}
+ n_3 \tilde{c} \tilde{d} + n_4 \tilde{d} b^{(2)}
\quad \too \quad  n_2 \tilde{c} \tilde{m} + n_3 \tilde{c} \tilde{d} + n_4 \tilde{d} b^{(2)}
\eeq

\begin{equation}
\centering
\begin{tabular}{|c||c|c||c|c|c|}\hline
 Field & \SU{F} & \SU{F+M-N} & \SU{F} &\SU{M+F}& $\SU{N}$ \\
\hline\hline
$b^{(3)}$ & $\Yfund$ & $\overline{\Yfund}$ & $ {\bf 1}$
  & $ {\bf 1}$ & $ {\bf 1}\vbr$ \\\hline
$\tilde{c}$ & $\overline{\Yfund}$ & {\bf 1} & {\bf 1} &
 $ \overline{\Yfund}$ 
  & $ {\bf 1}\vbr$ \\ \hline
$\tilde{d}$ & \Yfund & $ {\bf 1}$ & $\overline{\Yfund}$ & 
$ {\bf 1}$ & $ {\bf 1}\vbr$ \\
  \hline
$\tilde{m}$ & \Yfund & $ {\bf 1}$ & {\bf 1} & $ {\bf 1}$ & 
  $\overline{\Yfund} \vbr$ \\ \hline
$n_2$ & ${\bf 1}$ &  ${\bf 1}$ & ${\bf 1}$ & $ \Yfund$ & 
  $\Yfund \vbr$ \\ \hline
$n_3$ & ${\bf 1}$ & $ {\bf 1}$ & $\Yfund$ & $ \Yfund$ & 
 $ {\bf 1}  \vbr$ \\ \hline
$\tilde{n_4}$ & ${\bf 1}$ & $\Yfund$  & $\overline{\Yfund}$  & $ {\bf 1}$ & 
  ${\bf 1} \vbr$ \\ \hline
$p$ & $\overline{\Yfund}$ & ${\bf 1}$  & $\Yfund$  & $ {\bf 1}$ & 
  ${\bf 1} \vbr$ \\ \hline
\end{tabular}
\end{equation}
\beq
W = p \tilde{d} + n_2 \tilde{c} \tilde{m}
+ n_3 \tilde{c} \tilde{d} + p b^{(3)}\tilde{n_4}
\quad \too \quad n_2 \tilde{c} \tilde{m} + n_3 \tilde{c} b^{(3)} \tilde{n}_4
\eeq \vspace{-1cm}

\begin{equation}
\centering
\begin{tabular}{|c||c|c||c|c|c|}\hline
 Field & \SU{M} & \SU{F+M-N} & \SU{F} &\SU{M+F}& $\SU{N}$ \\
\hline\hline
$b^{(4)}$ & $\Yfund$ & $\Yfund$ &
  $ {\bf 1}$
  & $ {\bf 1}$ & $ {\bf 1}\vbr$ \\\hline
$c^{(2)}$ & $\overline{\Yfund}$ & {\bf 1} & {\bf 1} &
 $ \Yfund$ 
  & $ {\bf 1}\vbr$ \\ \hline
$m^{(2)}$ & \Yfund & $ {\bf 1}$ & {\bf 1} & $ {\bf 1}$ & 
  $\Yfund \vbr$ \\ \hline
$n_2$ & ${\bf 1}$ &  ${\bf 1}$ & ${\bf 1}$ & $ \Yfund$ & 
  $\Yfund \vbr$ \\ \hline
$n_3$ & ${\bf 1}$ & $ {\bf 1}$ & $\Yfund$ & $ \Yfund$ & 
 $ {\bf 1}  \vbr$ \\ \hline
$\tilde{n_4}$ & ${\bf 1}$ & $\Yfund$  & $\overline{\Yfund}$  & $ {\bf 1}$ & 
  ${\bf 1} \vbr$ \\ \hline
$q_1$ & ${\bf 1}$ & ${\bf 1}$  & ${\bf 1}$  & $ \overline{\Yfund} $ & 
  $\overline{\Yfund} \vbr$ \\ \hline
$q_2$ & ${\bf 1}$ & $\overline{\Yfund} $ & ${\bf 1}$ & $\overline{ \Yfund}$ & 
 $ {\bf 1}  \vbr$ \\ \hline
\end{tabular}
\end{equation}
\beq 
W = q_1 n_2 + q_2 n_3 \tilde{n}_4
+ q_1 c^{(2)} m^{(2)} + q_2 b^{(4)} c^{(2)} 
\quad \too \quad  q_2 n_3 \tilde{n}_4 + q_2 b^{(4)} c^{(2)} 
\eeq \vspace{-0.8cm}

\begin{equation}
\centering
\begin{tabular}{|c||c|c||c|c|c|}\hline
 Field & \SU{M} & \SU{N} & \SU{F} &\SU{M+F}& $\SU{N}$ \\
\hline\hline
$b^{(5)}$ & $\overline{\Yfund}$ & $\Yfund$ &
  $ {\bf 1}$
  & $ {\bf 1}$ & $ {\bf 1}\vbr$ \\\hline
$c^{(2)}$ & $\overline{\Yfund}$ & {\bf 1} & {\bf 1} &
 $ \Yfund$ 
  & $ {\bf 1}\vbr$ \\ \hline
$m^{(2)}$ & \Yfund & $ {\bf 1}$ & {\bf 1} & $ {\bf 1}$ & 
  $\Yfund \vbr$ \\ \hline
$n_3$ & ${\bf 1}$ & $ {\bf 1}$ & $\Yfund$ & $ \Yfund$ & 
 $ {\bf 1}  \vbr$ \\ \hline
$n_4^{(2)}$ & ${\bf 1}$ & $\Yfund$  & $\Yfund$  & $ {\bf 1}$ & 
  ${\bf 1} \vbr$ \\ \hline
$\tilde{q_2}$ & ${\bf 1}$ & $\overline{\Yfund} $ & ${\bf 1}$ &$\Yfund$ & 
 $ {\bf 1}  \vbr$ \\ \hline
$r_1$ & ${\bf 1}$ & ${\bf 1}$  & $\overline{\Yfund}$  &
   $\overline{\Yfund}$ & 
  ${\bf 1} \vbr$ \\ \hline
$r_2$ & $\Yfund$ & ${\bf 1} $ & ${\bf 1}$ &$\overline{\Yfund}$ & 
 $ {\bf 1}  \vbr$ \\ \hline
\end{tabular}
\end{equation}
\begin{equation} 
W = r_1 n_3 + r_2 c^{(2)} + r_1 n_4^{(2)} \tilde{q}_2 + r_2 b^{(5)}
\tilde{q}_2 \quad \too \quad 0
\end{equation}


\end{document}